\documentclass[11pt,twoside]{article}

\usepackage{asp2014}

\aspSuppressVolSlug
\resetcounters

\begin{document}

\title{IAU WG, Data-driven Astronomy Education and Public Outreach, current status and working plans}

\author{Chenzhou~Cui,$^1$ Shanshan~Li,$^1$}

\affil{$^1$National Astronomical Observatories, Chinese Academy of Sciences (CAS), 20A Datun Road, Beijing 100012, China; \email{ccz@nao.cas.cn}}

% This section is for ADS Processing.  There must be one line per author.

\paperauthor{Chenzhou~Cui}{ccz@nao.cas.cn}{}{National Astronomical Observatories, Chinese Academy of Sciences (CAS)}{Chinese Virtual Observatory}{Beijing}{}{100012}{China}
\paperauthor{Shanshan~Li}{lishanshan@nao.cas.cn}{}{National Astronomical Observatories, Chinese Academy of Sciences (CAS)}{Chinese Virtual Observatory}{Beijing}{}{100012}{China}

\begin{abstract}
IAU Inter-Commission B2-C1-C2 WG Data-driven Astronomy Education and Public Outreach (DAEPO) was launched officially in April 2017. With the development of many mega-science astronomical projects, for example CTA, DESI, EUCLID, FAST, GAIA, JWST, LAMOST, LSST, SDSS, SKA, and large scale simulations, astronomy has become a Big Data science. Astronomical data is not only necessary resource for scientific research, but also very valuable resource for education and public outreach (EPO), especially in the era of Internet and Cloud Computing. IAU WG Data-driven Astronomy Education and Public Outreach is hosted at the IAU Division B (Facilities, Technologies and Data Science) Commission B2 (Data and Documentation), and organized jointly with Commission C1 (Astronomy Education and Development), Commission C2 (Communicating Astronomy with the Public), Office of Astronomy for Development (OAD), Office for Astronomy Outreach (OAO) and several other non IAU communities, including IVOA Education Interest Group, American Astronomical Society Worldwide Telescope Advisory Board, Zooniverse project and International Planetarium Society. The working group has the major objectives to: Act as a forum to discuss the value of astronomy data in EPO, the advantages and benefits of data driven EPO, and the challenges facing to data driven EPO; Provide guidelines, curriculum, data resources, tools, and e-infrastructure for data driven EPO; Provide best practices of data driven EPO. In the paper, backgrounds, current status and working plans in the future are introduced. More information about the WG is available at:
http://daepo.china-vo.org/
\end{abstract}

\section{Backgrounds}

Astronomy education and public outreach (EPO) are always important to not only astronomical community but also the entire society. For astronomical community, they will determine the quality and quantity of the students who are willing to study astronomy in college and to pursue a lifetime career in astronomy. By proper EPO activities, people will have more understanding and appreciation of astronomy and the general public will support astronomical research more strongly. For the society, basic astronomy knowledge as long as other basic scientific knowledge should be common sense. The understanding of the sun, the moon, the earth and the universe will enrich daily life and study of everyone.

While the necessity of astronomy EPO is obvious, most teachers and organizations who are engaged in astronomical EPO related work still work in the traditional ways. Limited astronomical images making the exhibitions and books rich and colorful but fail to notice that without the real data, misunderstand of the concept in Astronomy education can be easily happened. Educators need to spend much more time than they expected to explain some astronomy concept. Still, after all the time and efforts they put in, students and the public might not be able to fully understand what is the real universe looks like. The EPO of astronomy needs new ways urgently.

Astronomy along with many other disciplines and industries has been entered the Big Data era. With the operation of many mega-science astronomical projects, for example, Cherenkov Telescope Array(CTA), Dark Energy Spectroscopic Instrument(DESI), EUCLID telescope, Global Astrometric Interferometer for Astrophysics(GAIA), James Webb Space Telescope(JWST), Large Synoptic Survey Telescope(LSST), Sloan Digital Sky Survey(SDSS), Square Kilometer Array(SKA), Five hundred meter Aperture Spherical radio Telescope (FAST) and Large Sky Area Multi-Object Fiber Spectroscopic Telescope(LAMOST), tens of terabyte scientific data generated every day. Astronomy research becomes a data-driven scientific activity. Though some voice said that the flood-like data didn't really enrich our knowledge and the information we got is decreasing \citep{dunham2006data}, most people believe as the exponential growth of data volumes accumulated, scientists, especially astronomers, are offered numerous opportunities to make excitement discoveries, to refine the models they built before and to fill the blanks we have on our knowledge trees.\citep{2009astro2010P...7B} In addition to that, even more people realize the huge amount of astronomical data is valuable resource for education and public outreach. 

The Virtual Observatory (VO) aims to provide a research environment that will open up new possibilities for scientific research based on data discovery, efficient data access, and interoperability. It is envisioned as a complete, distributed research environment for astronomy with large and complex data sets, by federating geographically distributed data and computing assets, and the necessary tools and expertise for their use.\citep{Cui2008} Furthermore, Astroinformatics is a bridge between astronomy and ICT (Information and Computation Technology) and applied computer science. The motivation is to engage a broader community of researchers, both as contributors and as consumers of the new methodology for data-intensive astronomy, thus building on the data-grid foundations established by the VO framework.\citep{Longo2014}

Big Data in astronomy is challenging our traditional research approaches and will radically transform how we train the next generation of astronomers, whose experiences with data are now increasingly more virtual (through online databases) than physical (through trips to mountaintop observatories). Data-driven science education (not only in astronomy) is not only increasingly vital for the next scientists but also for the next generation public, through which students are trained to access large distributed data repositories, to conduct meaningful scientific inquiries into the data, to mine and analyze the data, and to make science discoveries.\citep{Borne2010}

\section{Data Driven Astronomy Education and Public Outreach (DAEPO)}

Astronomical data is not only necessary resource for scientific research, but also very valuable resource for education and public outreach, especially in the era of Internet and Cloud Computing. Very specifically target on the data-driven EPO of astronomy, the International Astronomical Union Inter-Commission Working Group Data-driven Astronomy Education and Public Outreach (DAEPO) was launched officially in April 2017. This working group is hosted at the IAU Division B (Facilities, Technologies and Data Science) Commission B2 (Data and Documentation), and organized jointly with Commission C1 (Astronomy Education and Development), Commission C2 (Communicating Astronomy with the Public), Office of Astronomy for Development (OAD), Office for Astronomy Outreach (OAO) and several other non IAU communities, including International Virtual Observatory Alliance (IVOA) Education Interest Group, American Astronomical Society Worldwide Telescope Advisory Board, Zooniverse project and International Planetarium Society (IPS).(Figure \ref{fig1})

\articlefigure[width=.55\textwidth]{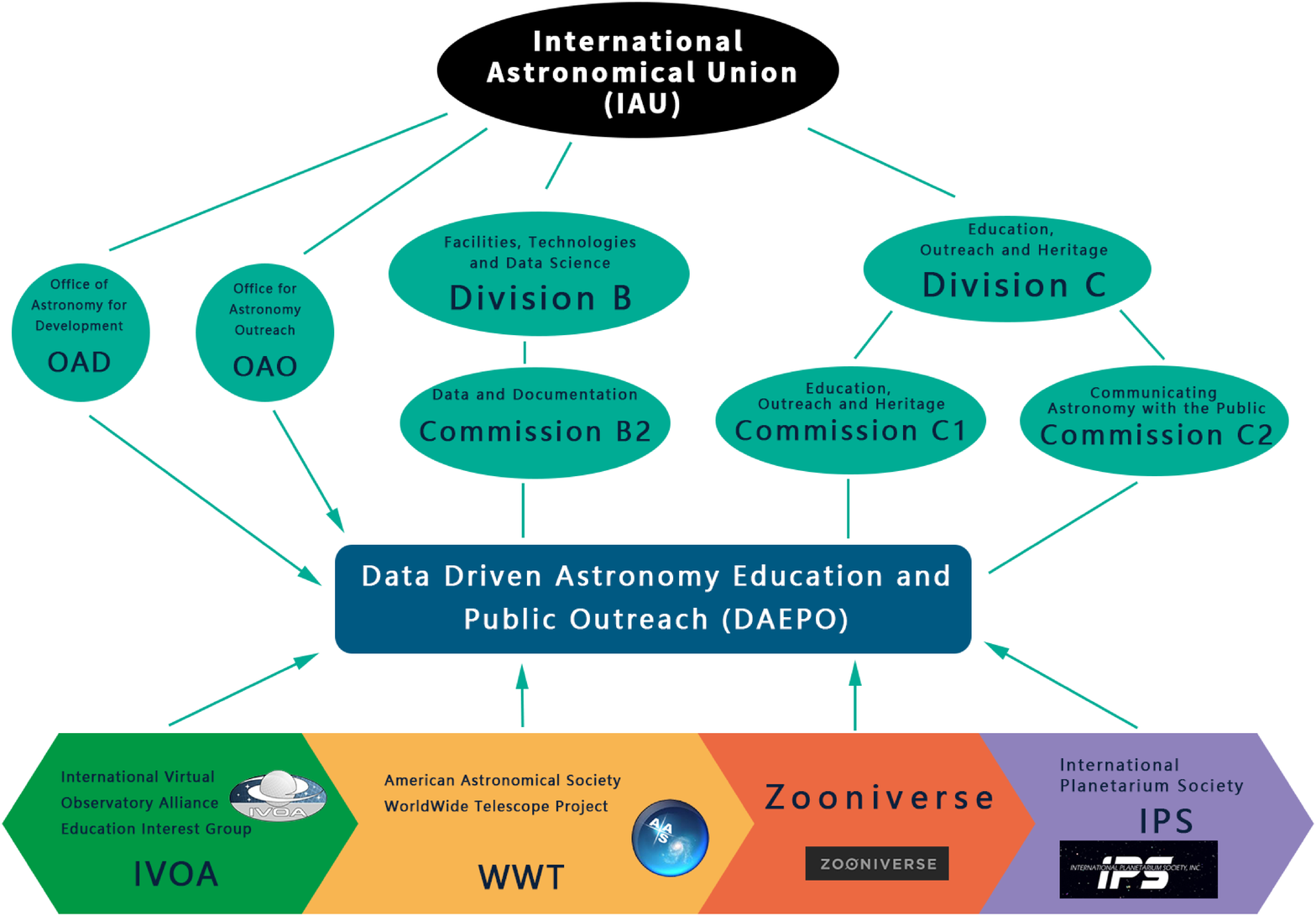}{fig1}{DAEPO Ecosystem}
\section{Current status}
During the very first year, most efforts of the WG are related to collect information about astronomy EPO, including current status of the astronomy EPO in different countries and regions, finding out if there is any possible environment and opportunities to carry out any form of data-driven astronomy EPO activities and practices. We keep following some of the very successful data related astronomy citizen science projects, like Worldwide Telescope (WWT) and ZOONIVERS. We also keep track of projects like Data to Dome from IPS and ESO, and many others.

In May 2017, Shanghai, Dr. Chenzhou Cui reviewed data-driven related astronomy EPO projects on IVOA Northern Spring Meeting, tried to spread out the idea of DAEPO and to inspire astronomers and engineers who working on or interested in astronomical data to pay close attention to the potential of the data using in education and public outreach areas. Led by Dr. Chenzhou Cui, Chinese Virtual Observatory (China-VO) is always interested in using astronomical data to improve astronomy education and public outreach. In 2015, a data-driven public outreach project called Popular Supernova Project (PSP) was initiated as the first citizen science project in Astronomy in China.\citep{ShanshanLi2016} A dozen of supernova or nova have been discovered by public users including a ten-year old primary school student.
 
On the ISE2A symposium at Utrecht in July 2017, another talk was given by Dr. Chenzhou Cui about IAU DAEPO WG. This is a symposium specifically concerned about astronomy education. This report introduced the idea of using astronomical data in science education and public outreach to astronomy educators around the world. 

As a new thinking and trend of education and public outreach, the concept of DEAPO is spread by WG members at different events, including ADASS 2017, Astroinformatics 2017, etc. 

\section{Working plans}
IAU DAEPO WG is ready to keep pushing forward the idea of bring real astronomical data into the classroom of primary school, middle school, and into universities or even to public places like planetarium and science museum. The OBJECTIVES of DAEPO working group are:

\begin{itemize}
\checklistitemize

\item 1. Act as a forum to discuss the value of astronomy data in EPO, the advantages and benefits of data driven EPO, and the challenges facing to data driven EPO.

\item 2.Provide guidelines, curriculum, data resources, tools, and e-infrastructure for data driven EPO.

\item 3.Provide best practices of data driven EPO.
\end{itemize}

In order to achieve these goals, the working group will carry out further information collection efforts in the form of questionnaires. At the same time, the WG continues involving propaganda activities on international meetings. On the IAU General Assembly in 2018 in Vienna, the working group will co-host a Focus Meeting: FM14 IAU's role on global astronomy outreach, the latest challenges and bridging different communities. Furthermore, the DAEPO WG is also involved in preparing for the IAU Astronomy education symposium 2019, the International Symposium on Education in Astronomy and Astrobiology 2019, and some other events.

This work is supported by National Natural Science Foundation of China (NSFC)(11503051, 61402325), the Joint Research Fund in Astronomy (U1531111, U1531115, U1531246, U1731125, U1731243) under cooperative agreement between the NSFC and Chinese Academy of Sciences (CAS). We would like to thank the National R\&D Infrastructure and Facility Development Program of China, "Earth System Science Data Sharing Platform" and "Fundamental Science Data Sharing Platform" (DKA2017-12-02-XX). Data resources are supported by Chinese Astronomical Data Center (CAsDC) and Chinese Virtual Observatory (China-VO). 

\bibliography{P6-3}
\end{document}